\input  vanilla.sty
\baselineskip 16pt
\magnification=\magstep1  
\TagsOnRight
\font\large=cmr10  scaled \magstep1
\font\Large=cmr10   scaled  \magstep2
\font\huge=cmr10   scaled \magstep3

\def\ref#1{$[#1]$}
\def\plusminus{{\pm}}

\def\comm#1,#2{{\left[#1 , #2\right]}}
\def\acomm#1,#2{{\left\lbrace #1,#2 \right\rbrace }}
\def\ttag#1{{\vskip -16pt \rightline{(#1)}}}

\def\Sum{{\sum}}

\def\pmb#1{\setbox0=\hbox{$#1$}%
                        \kern-.025em\copy0\kern-\wd0
                        \kern.05em\copy0\kern-\wd0
                        \kern-.025em\raise.0433em\box0}

\def\ref#1{$^{[#1]}$}

\def\ni{\noindent}
\def\Int{\int}

\footline={\hfil}
\vglue 2.0truecm
\setbox1=\hbox{$\bigotimes$}

\centerline {\huge Top and Higgs Masses in Dynamical Symmetry Breaking
\footnote{This manuscript has been authored under NSF supported
research Contract No. PHY91-13117 and DOE supported research
Contract No. DE-AC02-76CH00016.}}
\vglue 2.0truecm
\centerline{\Large David E. Kahana\footnote{e-mail: kahana\@scorpio.kent.edu}}
\vglue 10pt
\centerline{Center for Nuclear Research}
\centerline{Kent State University}
\centerline{Kent, OH 44242-0001}
\vskip 12pt
\centerline{\it and}
\vskip 12pt
\centerline
{\Large Sidney H. Kahana\footnote{e-mail: kahana\@bnlnth.phy.bnl.gov}}
\vglue 10pt
\centerline{Physics Department}
\centerline{Brookhaven National Laboratory}
\centerline{Upton, NY 11973}

\vglue 15truemm

\centerline {\underbar{\large ABSTRACT}}
A  model for composite electroweak bosons is re-examined to establish
approximate ranges for the initial predictions of the top and Higgs masses.
Higher order corrections to this $4$-fermion theory at a high mass scale where
the theory is matched to the Standard Model have little effect, as do wide
variations in this scale. However, including all one loop evolution and
defining the masses self-consistently, at their respective poles, moves the
top mass upward by some $10$ GeV to near $175$ GeV and the Higgs mass down by a
similar amount to near $125$ GeV.
 
\vfill\eject
\pageno=1
\footline={\hss\tenrm\folio\hss}
\def\ff{$4$-fermion }
\def\ref#1{[{\bf #1}]}
\leftline{\bf 1. Introduction}
\bigskip
\par
In this paper we refine predictions for the top and Higgs masses made in 
an earlier work on dynamical symmetry breaking \ref{1}. The specific
\ff model of dynamical symmetry breaking presented in Reference 1
(see the Lagrangian in Eq.1 below) may perhaps be ultimately  viewed as the 
low mass limit of a gauge theory at some very high scale 
$\Lambda$, with primordial
boson masses, $M_B \sim O(\Lambda)$. This scale then acts as an effective 
cutoff for the \ff theory. Certainly, no explanation is 
presented here for the 
number and character of elementary fermions in the modeling
nor for the large disparity in mass scales, i.e. $m_f\ll M_B$.
Rather, a central point of our calculation is that new, composite, bosons
with masses near $2 m_f$ arise naturally in the theory.
These are just fermion--antifermion bound states produced by the \ff
interaction. This phenomenon is well described in the papers of
Nambu and Jona--Lasinio on the four fermion theories \ref{2}, and has been
exploited by many authors \ref{3,4,5,6,7,8}. Since the scale $\Lambda$ at which
any new physics enters is so high, the theory is in fact a weak coupling,
albeit constrained, version of the Standard Model for scales well below 
$\Lambda$.  

Previously \ref{1} we  abstracted simple, asymptotic mass relationships
from the 
\ff theory, and used these as boundary conditions on the standard model
renormalisation group (RG) equations. This was done at a matching scale 
$\mu\sim M_{GUTS}$, where the electroweak (EW) sector can still be treated
as approximately independent of QCD ($SU(3)_c$). Values for the top and 
Higgs masses then followed from downward
evolution of the top-Higgs and Higgs-self couplings to scales near $m_W$,
assuming no intervening structure.

In the present work we show that modifications in these asymptotic 
relationships, due to higher order corrections in the \ff theory
at the upper scale $\mu$, have a considerably diminished effect on  $m_t$ and $m_H$ at their much lower scale. Also, large,
several orders of magnitude, changes in $\mu$  affect the top hardly at 
all and the Higgs only slightly. However, a more consistent handling of 
the RG evolution moves the prediction for $m_t$ from approximately $165$  
GeV to nearer $175$ GeV, while that for $m_H$ moves from $140$ GeV to about
$125$ GeV.

%More importantly, perhaps,
%we consider the values suggested by the theory for the electroweak couplings
%$\alpha_2$, $\alpha$, and for the Higgs--top Yukawa coupling $\kappa_t$,
%and find these consistent with a single cutoff near the Planck scale.
\bigskip
\leftline{\bf 2.The 4-Fermion Theory}
\medskip
\par
In Ref.1, we indicated that a \ff interaction including vector terms led to
rather low, well-determined masses for the top-quark and Higgs. The model is
defined by the Lagrangian:
$$
\align
{\cal L} =\bar\psi i(\gamma \cdot \partial) \psi & - {1\over 2}
  [(\bar\psi G_S \psi)^2-(\bar\psi G_S \tau\gamma_5\psi)^2]\\
 & - {1\over 2} G_B^2(\bar\psi\gamma_\mu Y \psi)^2-{1\over 2}
      G_W^2(\bar\psi\gamma_\mu \tau P_L\psi)^2\\
\endalign
$$
\ttag{1}
\noindent 
in which very specific vector interactions have been added to the usual
scalar and pseudoscalar terms of NJL. The field operator is $\psi = \lbrace
f_i \rbrace$, and the index $i$ runs over all fermions,
$i=\lbrace(t,b,\tau,\nu_{\tau}),(c,s,...),...\rbrace$. The scalar-coupling
matrix $G_s$ is taken diagonal and the dimensionful couplings are adjusted to
produce the known fermion masses dynamically; in practice only the top
acquires an appreciable mass. The model admits bound states corresponding
to the Higgs as well as the gauge bosons of the standard electroweak theory,
and is essentially equivalent to the Standard Model below some high mass
scale $\mu$. It is the vector terms in Eq.1 which ensure the existence of the
Higgs, $Z$, and $W$ as composites with masses of the order of $ m_t$, thus
naturally explaining why the Standard Model bosons and the top appear to have
about the same mass.

In re-examining the predictions for the top and Higgs we do not presume to
seek precise values for their masses, but rather attempt to determine the
latitude in masses present in the modeling. Such a study is especially timely
in light of the search for the top being carried out at FNAL \ref{9}.  The
apparent paucity of top events in the latest data suggests a high mass for
the top, certainly it now seems $m_t$ is greater than 120 GeV and possibly
considerably higher. Present analyses of LEP data \ref{10} with respect to EW
corrections, suggest $m_t = 166 \plusminus 30$ GeV.

As usual in NJL the necessary fine tuning of the scalar coupling is
accomplished by solving the scalar gap equation,  whence diagonalisation
of the scalar action yields the Higgs mass formula:
$$
m_H(\mu) = 2 m_t(\mu)(1 + O(g_t^2))
\tag{2}
$$
Fine tuning determinines the dimensionful scalar coupling in
terms of the cutoff $\Lambda$,   
$$
G_t^2 = \frac{1}{\Lambda^2 - m_t^2 \ln \left(\frac{\Lambda^2}{m_t^2}\right)}
\tag{3}
$$

Bound states also exist in the vector sector defined by Eq.1 corresponding to
the $W$, $Z$, and the photon. A similar fine tuning of the vector coupling is
required, but here with the added physical interpretation that the photon
mass should vanish \ref{1}. This latter constraint leads, at lowest order in
the electroweak and Yukawa couplings, to the mass relationship
$$
m_W^2(\mu) = \frac{3}{8} m_t^2(\mu) 
\tag{4}
$$
To the same order in couplings, the required diagonalision 
of the neutral vector boson
action results in
$$
\sin^2 (\theta_W) = (\Sum_i Q_i^2)^{-1}=\frac{3}{8},
\tag{5}
$$
with the denominator on the right hand side of Eq.(5) being summed over
the charges $Q_i$ in one generation.

The dimensionful couplings of the \ff theory are replaced, after fine-tuning
and wave function renormalisation, by the dimensionless couplings of the
Standard Model \ref{11,1}, and the gradient expansion of the effective action
is in fact an expansion in these dimensionless electroweak couplings. One has
for the scalars
$$
\align 
g_S &= G_S Z_S^{-\frac{1}{2}},\\
Z_S &= \frac{1}{2} Tr \left[G_S^2\frac{1}{(\partial^2 + M^2)^2}\right],\\
\endalign
$$
\ttag{6}
\ni where the fermion--scalar coupling matrix is  for the present
taken diagonal:
$$
(G_S)_{ij} = G_i \delta_{ij}.
\tag{7}
$$
 Similarly, for the vector couplings one has
$$ 
\frac{g_2}{2} = \frac{G_W}{\sqrt{Z_W}}
\qquad\text{and}\qquad  
\frac{g'}{2} = \frac{G_B}{\sqrt{Z_B}}
\tag{9}
$$
\ni and the usual relationship between $g_2$ and $g'$
$$
g_2 \sin (\theta_W) = g' \cos (\theta_W).
\tag{10}
$$  

From equations $\lbrace (2), (4), (5)\rbrace$, valid presumably at a scale
$\mu$ where the cross coupling between the EW and strong sectors is small but
still well below the cutoff $\Lambda$, we derived values for the top and
Higgs masses at a scale near $m_W$. The theory leading to these equations is
equivalent to the electroweak sector of the Standard Model below $\mu$, and
the framework for connecting the scales $\mu$ and $m_W$ is provided by the
Standard Model RG. Thus $SU(3)_c$ influences on the top and Higgs masses are
included through the renormalisation group, below the matching scale $\mu$.
\bigskip
\leftline{\bf 3. Renormalisation Group Evolution}
\medskip
\par
We turn now to the calculation of smaller effects, neglected in the initial
work, due to corrections in the \ff\ theory of higher order in the
electroweak couplings and to a more consistent treatment of the
evolution downward to experimental mass scales. Our basic equations are:
(1) the boundary condition relationships between the Higgs, top and W
masses including dependence on electroweak couplings and quark masses,
and (2) the RG evolution equations for the top-Higgs and Higgs-self couplings
$g_t$ and  $\lambda$. Defining \ref{12,13}
$$
\kappa_t = \frac{g_t^2}{2\pi},
$$
one has
\def\kt{\kappa_t}
\def\as{\alpha_S}
\def\aw{\alpha_W}
\def\aone{\alpha_1}
$$
\frac{d\kt}{dt} = \frac{9}{4\pi} \kt^2 - \frac{4}{\pi} \kt \as
- \frac{9}{8\pi} \kt \aw - \frac{17}{4\pi} \kt \aone,
\tag{12}
$$
with $\as$, $\aw$, $\aone$ taken equal to $\alpha_3$, $\alpha_2$, $\alpha_1$
respectively in reference \ref{12,13}, and $t=\ln(\frac{q}{m})$. We note that
with these choices 
$$
\align
m_t &= g_t v,\\
m_W &= \frac{g_W}{2} v,\\
\endalign
$$
\ttag{13}
\ni where $v$ is the standard EW vev.

Also taking $m_H^2 = 2 \lambda v^2$ the evolution equation for the Higgs
self-coupling is, to the same (one-loop) order \ref{14}:
\def\l{\lambda}
$$
\frac{d\l}{dt} = \frac{1}{16\pi^2} \left\lbrace 12 \l^2 + 6 \l g_t^2
- 3g_t^4 - \frac{3}{2} \l \left(3 g_W^2 + {g'}^2\right)
+ \frac{3}{16}
\left(2 g_W^4 + \left(g_W^2 + {g'}^2\right)^2\right)\right\rbrace .
\tag{14}
$$
Redefining the standard choice of couplings \ref{12}
$$  
\aone =\frac{5}{3} \alpha'
\qquad
\text{with}
\qquad \alpha_1 = \frac{g_1^2}{4\pi},
\qquad \alpha' = \frac{{g'}^2}{4\pi}
\tag{15}
$$
and setting
\def\s{\sigma}
$$
\sigma = \frac{\lambda}{4\pi}
\tag{16}
$$
results in
$$
\frac{d\s}{dt} = \frac{1}{2\pi} \left\lbrace 
12 \s^2 + 6 \s \kt
- 3\kt^2 
- \frac{9}{2} \s 
\left(\alpha_W + \frac{1}{5} \alpha_1\right)
+ \frac{3}{16} 
\left(2 \alpha_W^2 +
\left(\alpha_W + \frac{3}{5} \alpha_1 \right)^2
\right)
\right\rbrace.
\tag{17}
$$

Equations (2) and (4) impose boundary conditions on equations (12) and (17)
at the scale $\mu$. These are to lowest order
$$
m_t^2 = \frac{8}{3}\,m_W^2 
$$
and
$$
m_H^2 = 4 m_t^2 = \frac{32}{3} \,m_W^2,
\tag{18}
$$
which can be restated to include higher orders:
$$
\frac{\kt}{\alpha_2} (\mu) = \frac{4}{3} + O(g_i^2)
\qquad\text{and}\qquad
\frac{\s}{\alpha_2} (\mu) = \frac{4}{3} + O(g_i^2).
\tag{19}
$$
%We turn our attention first to the corrections implied in Eq(19). The gap
Such corrections can come from two sources, higher order $1/N$, multi-loop,
contributions to the effective action, and more trivial $1/ln(\Lambda)$
terms within the lowest order.
%equation generation of mass for the bosons is correct to leading order
%in $1/N$. Such corrections could in principle be done either in the NJL model,
%or in the context of an underlying gauge theory with vector bosons having
%masses on the order of the cutoff $\Lambda$. However such a calulation
%is somewhat involved. Since here $N=4 N_c=12$, and we demonstrate that
%the predictions for $m_t$ and $m_H$ are stable with respect to small changes
%at the scale $\mu$ we do not pursue $1/N$ corrections here.
\noindent The latter arise, for example, from the proper generalised form of
Eq.4:
$$
m_W^2 = \frac{1}{2}
\frac{\Sum_i\, m_i^2 
\left[ \ln
\left(\frac{\Lambda^2}{m_i^2}+1 
\right)-1 
\right]}
{\Sum_i \, \frac{r_i}{6} 
\left[ 
\ln
\left(
\frac{\Lambda^2}{m_i^2}+1 
\right)-\frac{11}{6} 
\right]},
\tag{20}
$$
where the sum is over all fermions and
$
r_i =\alpha^{-2}\left(\beta^4 (y_{Li}^2 + y_{Ri}^2)
- \beta^2 \alpha^2 y_{Li} \tau^3_i
+ \alpha^4 \tau^3_i \tau^3_i\right),
$
while $y_{Li}, y_{Ri}\text{ and } \tau^3_i$ are the fermion hypercharges and
isospins. Equation (4) is obtained from (20) by keeping only the top mass and
ignoring terms of order $(\ln (\Lambda))^{-1}$.  These terms are of higher
order in the electro weak couplings; for example the Higgs-top Yukawa
coupling is, from (6), proportional to $(\ln(\Lambda))^{-1}$.

We note parenthetically that the basic $SU(5)$ symmetry evident in Eqs(4,5)
results from the $\bar {\text{\bf 5}} + \text{\bf 10}$ generational structure
$(u,d,e_{L,R},\nu_{R})$
built into the present model, and follows from (20) in the limit of large
$\Lambda$. We also note that the $\frac{3}{8}$ appearing in the lowest order
(Eq(4)) for $m_W^2$ is more properly written
$$
\frac{m_W^2}{m_t^2} = \frac{3}{8} \frac{n_g}{n_c},
\tag{21}
$$
and so is not simply $\sin^2(\theta_W)$ but instead depends on the number
of colours as well as the number of massive fermion generations.  We find
that the several percent change implied in Eq(20) relative to Eq(4)
produces a considerably smaller change in $m_t$, less than one percent. Thus,
to the accuracy meaningful here, we can perhaps ignore these corrections as
well as other higher order $1/N$ effects arising from discarded, incoherent,
summations over fermions.
\bigskip
\leftline{\bf 4. Solution of the RG Equations.}
\medskip
\par
It is possible to obtain an explicit solution to Eq(12), and a perturbative
solution for Eq(17). For the top evolution one has,
making a simple transformation of Eq(12)
$$
\frac{d}{dt} \frac{1}{\kt} = - \frac{9}{4\pi}
+ \frac{1}{\kt} \left( \frac{4}{\pi} \as + \frac{9}{8\pi} \aw + 
\frac{17}{40\pi} \aone \right),
\tag{22}
$$
with the one parameter family of solutions
$$
\align
\frac{1}{\kt} &=
\left[
\frac{(1 + \alpha_{S0} b_S t)^{8/7} (1 + \alpha_{W0} b_W t)^{27/38}}
{(1 - \alpha_{10} b_1 t)^{17/82}}
\right]\\
&\hskip +3.0truecm
\times
\left[
D
-
\frac{9}{4\pi} \Int_{0}^{t} dt' 
\frac{(1 - \alpha_{10} b_1 t')^{17/82}} 
{(1 + \alpha_{S0} b_S t')^{8/7} (1 + \alpha_{W0} b_W t')^{27/38}}
\right]\\
\endalign
$$
\ttag{23}
\def\as0{\alpha_{S0}}
\def\aw0{\alpha_{W0}}
\def\a10{\alpha_{10}}
\ni Here $\as0$, $\aw0$, and $\a10$ are the couplings at
$t = \ln \frac{m_W}{m_W}=0$, and the constants
$b_S = \frac{7}{2\pi}$, $b_W = \frac{19}{12\pi}$ and $b_1 =\frac{41}{20\pi}$
determine the evolution of the $SU(3)$, $SU(2)$, and $U(1)$ couplings 
respectively. The constant $D$ in Eq(23) is given by
$$
D = \frac{1}{\kt(0)},
\tag{24}
$$
and directly yields the running top mass at the scale $m_W$ from
$$
m_t^2(m_W) = \frac{2\kt(0)}{\alpha_W (0)} m_W^2(m_W).
\tag{25}
$$
To self-consistently determine the physical top mass as a pole in the
top quark propagator, one must run $m_t(m_W)$ back up to get $m_t(m_t)$.

The cross coupling in Eq(17) complicates its 
solution. The pure scalar self-coupling result 
\def\so{\sigma_0}
$$
\so (t) = \frac{\so(0)}{1 - \frac{6}{\pi} \so (0) t},
\tag{26}
$$
may be improved  perturbatively
\def\s1{\sigma_1}
$$
\sigma (t) = \so (t) + \s1 (t) .
\tag{27}
$$
Linearising in the small correction $\s1 (t)$ produces
$$
\s1 (t) = e^{-v(t)} \Int_{t_\mu}^t dt' g(t') e^{v(t')},
\tag{28}  
$$
with
$$
v(t) = - \Int_{t_\mu}^t dt' f(t'),
\tag{29a}
$$
$$
f(t) = \frac{12}{\pi} \sigma_0 (t)+
\frac{3\kt(t)}{\pi}
- \frac{9}{4\pi}\left[ \alpha_2(t) + \frac{1}{5} \alpha_1(t)\right],
\tag{29b}
$$
and
$$
g(t) = 
\frac{3}{\pi} \sigma_0(t) \kt(t)
- \frac{3\kt^2}{2\pi}
+ \frac{3}{32\pi}
\left[
2 \alpha_2^2(t) + \left( \alpha_2(t) + \frac{3}{5} \alpha_1(t) \right)^2
\right].
\tag{29c}
$$
Boundary conditions are introduced at $t_{\mu}= \ln\frac{\mu}{m_W}$ through
$$
\s1(t_\mu) = 0, \qquad \so(t_\mu) = \kt (t_\mu) = \frac{4}{3} \alpha_2(t_\mu)
+O(\alpha_i^2).
\tag{30} 
$$
Since  $\s1(t)$ is small over the range $m_H$ to $\mu$ (see Fig.1) there is
no need to include higher orders. 

Results from numerical integration of equations (23) and (28,29) are
displayed in Table 1, and Figs 1-4. We have varied the inputs to these
calculations, the strong and electroweak couplings $\alpha_{i0}$, $i =
1,W,S$ over a reasonable range, somewhat wider than the flexibility allowed
by present experiments. The W mass is fixed at 80.1 GeV.  There are no free
parameters in the theory, the couplings and $m_W$ being determined from
experiment. A possible exception is the cutoff $\Lambda$, which is surely
well above $\mu$ and has essentially no effect on $m_t$ and $m_H$. Any
dependence other than logarithmic on $\Lambda$ has been eliminated by fine
tuning, while residual $\ln (\Lambda)$ presence is transmuted into
dependence on the dimensionless couplings.

The effect of imposing boundary conditions sharply at a scale $\mu$ remains
to be examined. As we noted above, $\mu$ is that point, when one is evolving
downward in mass, at which the $g_i$ become interdependent.  For example, the
top quark evolution is strongly influenced by $SU(3)_c$ from $\mu\sim 10^{14}
$ downward, and the running of $\alpha_W$ is also significant.  Varying $\mu$
over four orders of magnitude from $\mu = 10^{10}$ GeV to $\mu = 10^{14}$ GeV
has practically no effect on $m_t$, and only a small effect on $m_H$. This
remarkable result is demonstrated in Fig.(2) for central choices of the
couplings, and lends credence to our use of a sharp boundary condition.

The one physical parameter sensitive to $\mu$ is the weak mixing angle
$\theta_W$. We indicated \ref{1} that, for one loop evolution,
$\sin^2(\theta_W)$ achieves its experimental value $\sim 0.23$ (at $m_W$) for
$\mu\sim 10^{13}$ GeV. Unlike GUTS, the present theory need not have a single
scale at which the gauge couplings are equal. The unification present in this
model simply implies that the Standard Model should evolve smoothly into the
effective \ff theory where the couplings become weak.  Table 1 displays the
value of the couplings at scale $\mu$; the $\alpha_i$ are the experimental
values determined at $m_W$ evolved upward to $\mu$ at 1-loop and $\kt(\mu)$
is obtained from the boundary condition $\frac{\kt}{\alpha_2} = \frac{4}{3}$.
It is clear that the couplings are indeed all small at $\mu$, again
justifying the placing of the boundary conditions there.

Figures (3) and (4) show the variations of $m_t$ and $m_H$ with the strong
and electroweak couplings, respectively. The strong coupling is less well
known. Using as central values $\alpha_{S0} = 0.107$, $\alpha_{W0}=0.0344$,
and $\alpha_{10} = 0.0169$ \ref{10,16}, we get $m_t\simeq 175$ GeV and
$m_H\simeq 125$ GeV. Included in the $175$ Gev is is a $6$ Gev reduction from
evolving the top self-consistently to its proper mass at $q = m_t$; for the
Higgs this effect is much smaller. Further small contributions to Eq(19), from
non-leading log terms in defining the top pole and from running the W mass,
more or less cancel. It is clear from the figures that $m_H$ is somewhat more
sensitive to all these changes, and so the remaining uncertainty in the mass
$125$ GeV is larger. This uncertainty nevertheless may be usefully bounded by
noting \ref{1} that a rather large arbitrary variation in the boundary condition
ratio $m_H/m_t$ from $2$ to $\sqrt{8}$ produces $\leq 15$ GeV change in
$m_H$.  One must also keep in mind that the top is confined and its mass
therefore subject to some ambiguity in definition.
\bigskip
\leftline{\bf 4. Conclusions.}
\medskip
In summary, one gets remarkably stable predictions for the top and Higgs
masses and in a parameter free fashion. The only inputs were the
experimentally known couplings and the W-mass. A characteristic prediction of
this type of theory is $m_h<m_t$, so that the Higgs, which is practically a
$t\bar t$ condensate, is deeply bound.

In view of the present dearth of events from the FNAL experiments with D\O\
and CDF, the above prediction for the top (near 175 GeV) may not be wholly
wild. In light of the recent unfortunate developments at the SSC, the
somewhat low prediction for the Higgs mass, near 125 GeV, may take
considerably longer to test.

Finally, there is the question of the number of generations. In Ref(1), we
indicated that a fourth generation, with massive quarks $m_{t'}\sim m_{b'}
\sim m_t$, implies a top mass near $115$ GeV. Such a constraint arises from the
sum rule (Eq(19)) for $m_W^2$. Present data at FNAL appear to rule out this
possibility.

\vfill\eject
\bigskip
\centerline{\bf Acknowledgements}
\medskip
One of the authors (SHK) would like to thank the Alexander von Humboldt
Foundation (Bonn, Germany) for partial support and Professor W. Greiner of the 
Johan Wolfgang Goethe University, Frankfurt, Germany for his hospitality.
\vfill\eject
\footline={\hfil}
\noindent{\bf References \hfil}
\vglue 0.4cm
\item{1.} 
{D.E.~Kahana and S.H.~Kahana, {\it Phys. Rev.} {\bf D43}, 2361 (1991)}
\item{2.}
{Y.~Nambu and G.~Jona--Lasinio, {\it Phys. Rev.} {\bf 122},
345 (1961)}
\item{3.}
{J.D.~Bjorken, {\it Ann. Phys.} (NY) {\bf 24}, 174 (1963)}
\item{4.}
{W.~Bardeen, C.T.~Hill, and M.~Lindner, {\it Phys. Rev.}
{\bf D41}, 1647 (1990)}
\item{5.}
{H.~Terazawa, Y.~Chikashige, and K.~Akama, {\it Phys. Rev.}
{\bf D15}, 480 (1977)}
\item{6.}
{T.~Eguchi, {\it Phys. Rev.} {\bf D14}, 2755 (1976)}
\item{7.}
{M.~Bando, T.~Kugo, and K.~Yanawaki, {\it Phys. Rep.} {\bf 164},
210 (1988)}
\item{\ }{M.~Suzuki, {\it Phys. Rev.} {\bf D37}, 210 (1988)}
\item{8.}{V.A.~Miranski, M.~Tanabashi, and K.~Yamawaki, {\it Mod. Phys. Lett.}
{\bf A4}, 1043 (1989)}
\item{9.}{Avi~Yagil, {\it et al.}, CDF Collaboration, ``Proceedings of the
$7$th Meeting of the American Physical Society, Division of Particles and
Fields, 10-14 November, 1992.'', Vol.\,1. Edited by Carl~H.~Albright,
Peter~H.~Kasper, Rajendran Raja and John Yoh; World Scientific (1993),
(and other contributions therein).}
\item{\ }{Ronald~J.~Mahas, {\it et al.}, D\O\ Collaboration,
``Proceedings of the $7$th Meeting of the American Physical Society, Division
of Particles and Fields, 10-14 November, 1992.'', Vol.\,1. Edited by
Carl~H.~Albright, Peter~H.~Kasper, Rajendran Raja and John Yoh;
World Scientific (1993), (and other contributions therein).}
\item{\ }{D\O\ Collaboration, in the ``Proceedings of the Lepton Photon
Conference, August 4-7, 1993, Cornell University, Ithaca, N.Y.}
\item{10.}{Samuel~Ting, Summary of LEP Results, ``Proceedings of the
$7$th Meeting of the American Physical Society, Division of Particles and
Fields, 10-14 November, 1992.'', Vol.\,1. Edited by Carl~H.~Albright,
Peter~H.~Kasper, Rajendran Raja and John Yoh; World Scientific (1993),
(and other contributions therein).}
\item{\ }{LEP Summary, in the ``Proceedings of the Lepton Photon
Conference, August 4-7, 1993, Cornell University, Ithaca, N.Y.}
\item{11.}
{G.~S.~Guralnik, K.~Tamvakis, {\it Nucl. Phys.} {\bf B148}, 283 (1979)}
\item{12.}
{W. Marciano, {\it Phys. Rev. Lett.} {\bf 62}, 2793 (1989)}
\item{13.}
{W.~Marciano, {\it Phys. Rev.} {\bf D41}, 219 (1990)}
\item{\ }
{W.~Marciano and A.~Sirlin, {\it Phys. Rev.} {\bf D22}, 2695 (1980)}
\item{14.}
{John F. Gunion, Howard E. Haber, Gordon Kane and Sally Dawson,
``The Higgs Hunter's Guide'', Addison-Wesley, New York (1990)}
\item{15.}
{H.~Georgi and S.~Glashow, {\it Phys. Rev. Lett.} {\bf 32}, 438 (1974)}
\item{\ }
{H.~Georgi, H.~Quinn and S.~Weinberg, {\it ibid.} {\bf 33}, 451 (1974)}
\item{16.}
{U.~Amaldi {\it et al.}, {\it Phys. Rev.} {\bf D36}, 1385 (1984)}
\item{\ }{and LEP in DPF92}
\item{17.}
{S.~Fanchiotti and A.~Sirlin, {\it Phys. Rev.} {\bf D41}, 319 (1990)}
\vfill\eject
\centerline{\bf Figure Captions}
\bigskip
\footline={\hfil}
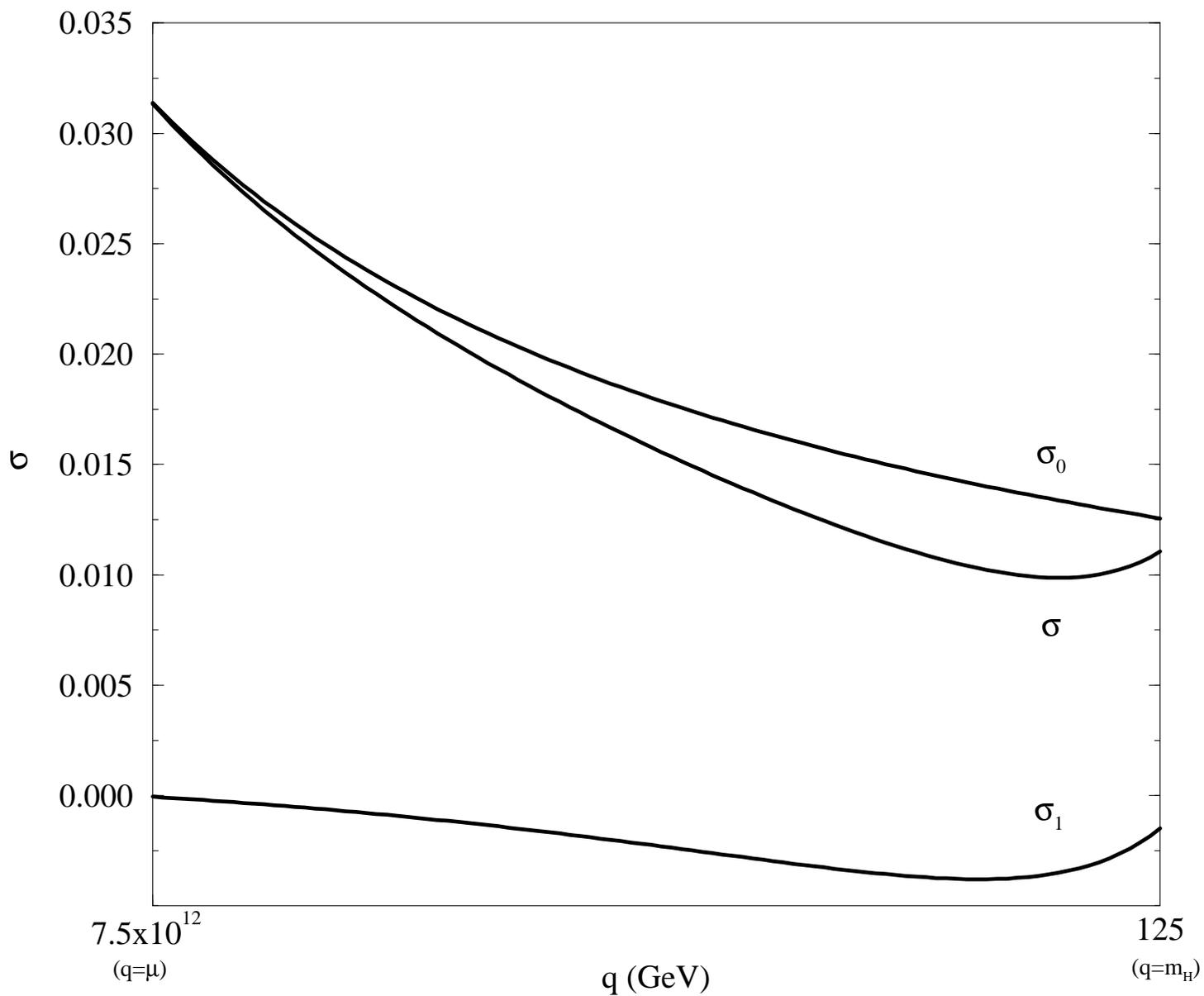
\item{\bf Fig. 1}
{Evolution of the reduced Higgs Self Coupling $\sigma = 
\sigma_0+\sigma_1$ over the range 
from $m_W$ to $\mu=10^{14}$. The perturbation $\sigma_1$ remains small.}
\smallskip
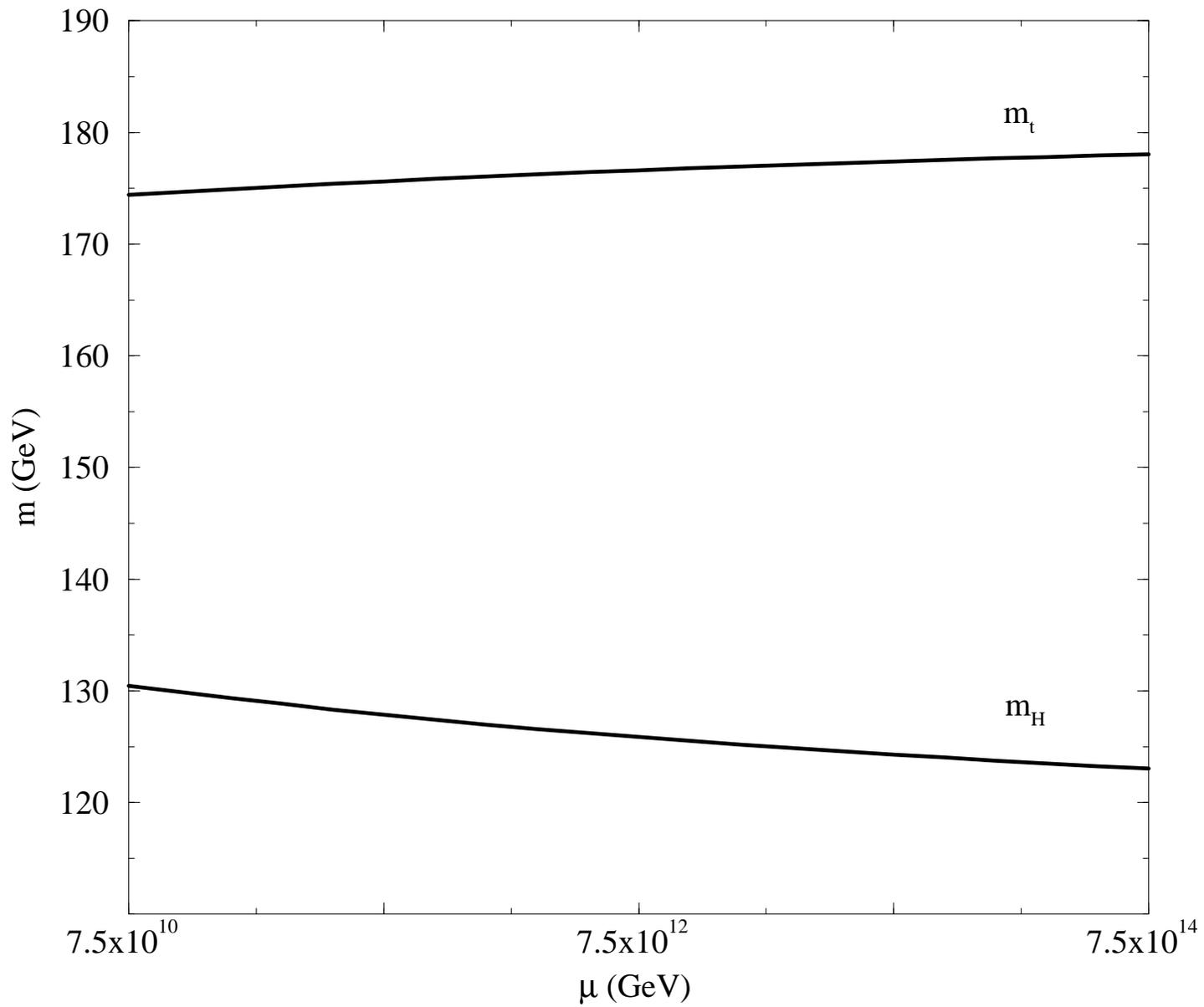
\item{\bf Fig. 2}
{Variation of the top and Higgs masses with the matching scale $\mu$ over
a range from $10^{10}$ to $10^{14}$ GeV. The scale $\mu=7.5\times 10^{12}$,
for which $\sin^2(\theta(\mu))=\frac{3}{8}$, is defined as a
`central value'.}
\smallskip
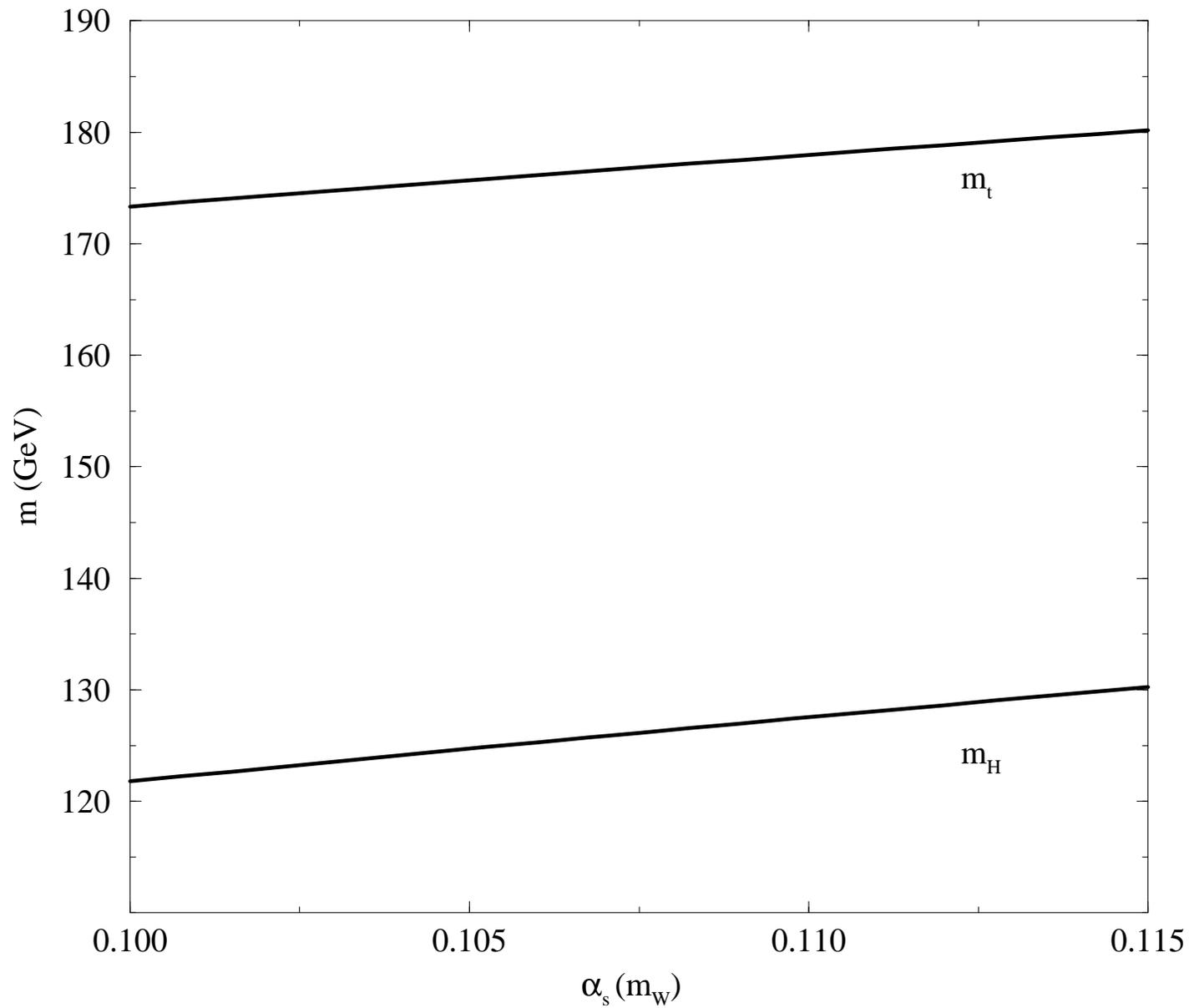
\item{\bf Fig. 3}
{Variation of $m_t$ and $m_H$ with the strong coupling; $\alpha_S=0.107$ is
considered the central value.}
\smallskip
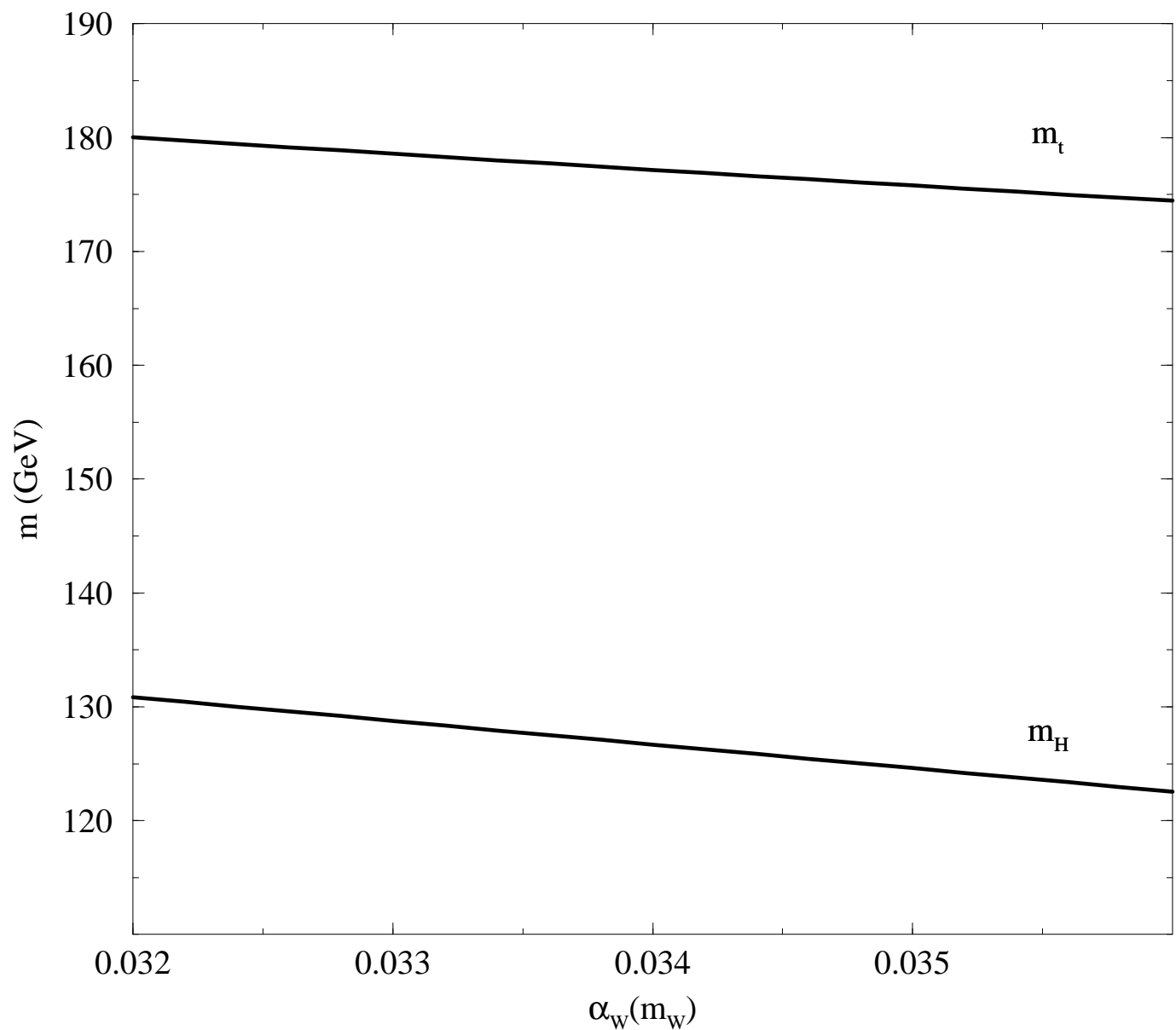
\item{\bf Fig. 4}
{Variation of $m_t$ and $m_H$ with the weak coupling;  $\alpha_W=0.0344$
is the central value.}
\vfill\supereject
\bye